**ORIGINAL ARTICLE**

**Open Access**

# Photon-counting computed tomography thermometry via material decomposition and machine learning

Nathan Wang[1*], Mengzhou Li[2] and Petteri Haverinen[3]

## Abstract

Thermal ablation procedures, such as high intensity focused ultrasound and radiofrequency ablation, are often used to eliminate tumors by minimally invasively heating a focal region. For this task, real-time 3D temperature visualization is key to target the diseased tissues while minimizing damage to the surroundings. Current computed tomography (CT) thermometry is based on energy-integrated CT, tissue-specific experimental data, and linear relationships between attenuation and temperature. In this paper, we develop a novel approach using photon-counting CT for material decomposition and a neural network to predict temperature based on thermal characteristics of base materials and spectral tomographic measurements of a volume of interest. In our feasibility study, distilled water, 50 mmol/L $CaCl_2$, and 600 mmol/L $CaCl_2$ are chosen as the base materials. Their attenuations are measured in four discrete energy bins at various temperatures. The neural network trained on the experimental data achieves a mean absolute error of 3.97 °C and 1.80 °C on 300 mmol/L $CaCl_2$ and a milk-based protein shake respectively. These experimental results indicate that our approach is promising for handling non-linear thermal properties for materials that are similar or dissimilar to our base materials.

## Introduction

Annually, over 100000 patients undergo thermal ablation procedures for a wide range of benign and malignant tumors [1]. As a primary example, high intensity focused ultrasound (US), which heats a focal region using a concave transducer, is an effective non-invasive treatment for prostate and other cancers [2, 3]. Currently, the delivery of the thermal dose is guided by invasive thermistors which can be fragile and only report temperatures from a limited number of points [4, 5]. Over the past decades, significant research efforts were devoted to extracting and analyzing thermal data from medical imaging modalities like US, magnetic resonance imaging (MRI), and computed tomography (CT). Among these modalities, CT is particularly advantageous for its real-time acquisition, high spatial resolution, and full-body coverage. In contrast, MRI has significant drawbacks in scanning speed, geometric accuracy, and cost, while US suffers from strong artifacts and restricted penetration through hard tissues and across air-tissue interfaces [6, 7].

While ionizing radiation to the patient is the main problem associated with CT, solutions are being rapidly developed over the past years. For instance, interior tomography allows for targeted imaging of a region of interest [8]. Also, data-driven methods

*Correspondence:
Nathan Wang
swang279@jhu.edu
[1] Department of Biomedical Engineering, Johns Hopkins University, Baltimore, MD 21218, USA
[2] Department of Biomedical Engineering, Rensselaer Polytechnic Institute, Troy, NY 12180, USA
[3] Aalto Design Factory, Aalto University, Espoo 02150, Finland

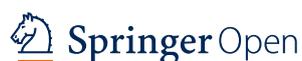

(i.e., machine learning and deep learning) have been applied to low-dose image reconstruction and denoising [9]. Synergistically, hardware-based innovations enabled photon counting CT (PCCT), which is a new frontier of medical imaging. PCCT can reduce radiation dose by eliminating electron noise, minimizing sensitivity to beam hardening through optimal X-ray photon weighting, increasing spatial resolution with fine detector pitch, and performing multiple material decomposition beyond the capabilities of dual energy CT [10, 11]. With FDA approval, these advancements have already been used in multiple clinical applications.

The ability for CT to measure temperature changes is based on the induced change in X-ray linear attenuation coefficient (LAC) as the result of thermal expansion. In general, heat applied to a tissue causes an increment in volume and thus decrement in density, which is observed as a drop in the LAC. The relationship between *CT* number, which is a normalized measure of the LAC expressed in Hounsfield units (HU), and temperature is modeled as Eq. 1.

$$\Delta CT(T) \approx -[1000 + CT(T_0)]\alpha \Delta T \qquad (1)$$

where $T_0$ is an initial baseline temperature, and $\alpha$ is the material-specific thermal expansion coefficient [12]. The change in HU per degree Celsius is called the thermal sensitivity and is often approximated as a constant over the relevant temperature range (approximately 30 °C to approximately 90 °C). This linear trend is confirmed in the prior studies which examined substances including water, fat, liver, kidney, etc. [13, 14]. Overall, studies have shown that CT thermometry can reach an impressive accuracy of 3-5 °C, but only after calibration to a given material [1]. While the principle of CT thermometry is conceptually simple, the variability in thermal sensitivity between different tissues, different patients, and under different scanning protocols is a critical challenge [1]. It would be difficult or impossible to obtain these highly specific measures in vivo, and clearly there are substantial differences between in vivo and ex vivo measurements because of the different physiological conditions. Furthermore, exposure to intense heat during thermal ablation may alter the thermal properties of the target region, introducing additional errors.

To address these significant problems with CT thermometry, here we present the first approach for PCCT thermometry that allows for superior material decomposition and data-driven temperature mapping relying on basis material data that do not need patient-specific calibration. Using PCCT to simultaneously capture the LAC of a substance at several energy levels, we can perform material decomposition, which is demonstrated in Eq. 2 for three base materials without loss of generality [15, 16].

$$\begin{cases} \mu(E_1) = V_1\mu_1(E_1) + V_2\mu_2(E_1) + V_3\mu_3(E_1) \\ \mu(E_2) = V_1\mu_1(E_2) + V_2\mu_2(E_2) + V_3\mu_3(E_3) \\ \qquad\quad 1 = V_1 + V_2 + V_3 \end{cases} \qquad (2)$$

$\mu_1$, $\mu_2$, and $\mu_3$ are the known energy-dependent LACs of the bases and $V_1$, $V_2$, and $V_3$ are the corresponding unknown volume fractions. Physically speaking, the LAC of a mixture of the base materials must be the linear combination of the LACs of the components with the corresponding volume fractions as the weighting factors.

Given the above, one might reasonably expect that thermal sensitivity could be linearly computed according to the material composition. In other words, given that $\mu_i(T) \approx \alpha_i(T - T_0) + \mu_i(T_0)$ where $T_0$ is a reference temperature and $\alpha_i$ is the thermal sensitivity, a linear model for the LAC for $n$ base materials would be as follows:

$$\mu(T) = \sum_{i=1}^{n} V_i\mu_i(T) = \alpha'(T - T_0) + \beta' \qquad (3)$$

where $\alpha' = \sum_{i=1}^{n} V_i\alpha_i$ and $\beta' = \sum_{i=1}^{n} V_i\mu_i(T_0)$ are the volume fraction weighted thermal sensitivity and offset respectively. In reality, thermal sensitivity relies primarily on thermal expansion, which is directly related to the strength of intermolecular bonds. Hence, the above linear model is generally inaccurate. Indeed, our experimental data presented in Fig. 2g shows that thermal sensitivity follows a quadratic/higher order relationship with the concentration of $CaCl_2$, indicating that data-driven modelling is suitable for PCCT thermometry. Since a fully connected neural networks with proper activations can approximate any continuous function, it is an ideal choice for non-linear prediction of temperature given spectrally resolved LAC values, which is essentially a multivariate regression task.

## Methods

In our feasibility study, we selected (1) water and aqueous solutions of (2) 50 mmol/L $CaCl_2$ and (3) 600 mmol/L $CaCl_2$ as our three base materials since the human body is characteristically composed of water and bone. These substances were heated in a hot water bath with precision temperature control and immediately transferred to a custom-built rectangular cuboid phantom with a digital thermometer (DS18B20 thermometer, ± 0.25 °C) shown in Fig. 1a. The thermal expansion of the acrylic phantom container is negligible in comparison to the substances being measured. The LAC values of the homogeneous base substances were measured in four energy bins (8-33 keV, 33-45 keV, 45-60 keV, and 60-100 keV) during transient cooling and at approximately every 5 °C

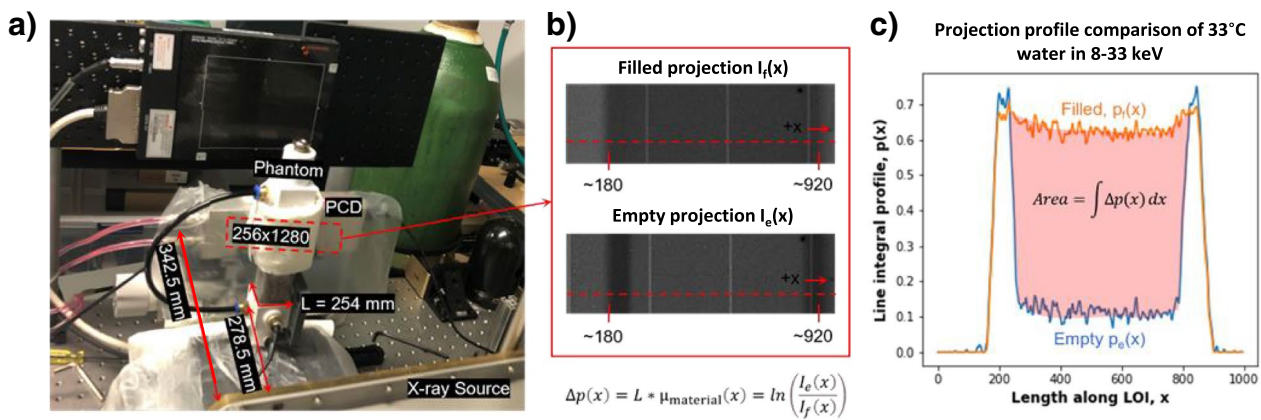

**Fig. 1** Illustration of the experimental setup and procedure. **a** Photo of the photon counting CT configuration used to take 2D projections (256 × 1280 pixels) of the phantom; **b** the 200th row in the projection is selected as the line of interest (LOI) and used to obtain the difference between the projection profiles of the phantom when it is filled and when it is empty. The projections have been contrast enhanced for better viewing and the vertical white lines corresponding to gaps between detector chips are removed during processing; **c** The difference in area between the projection profiles of the empty and filled phantom are used to determine the LAC of the liquid material

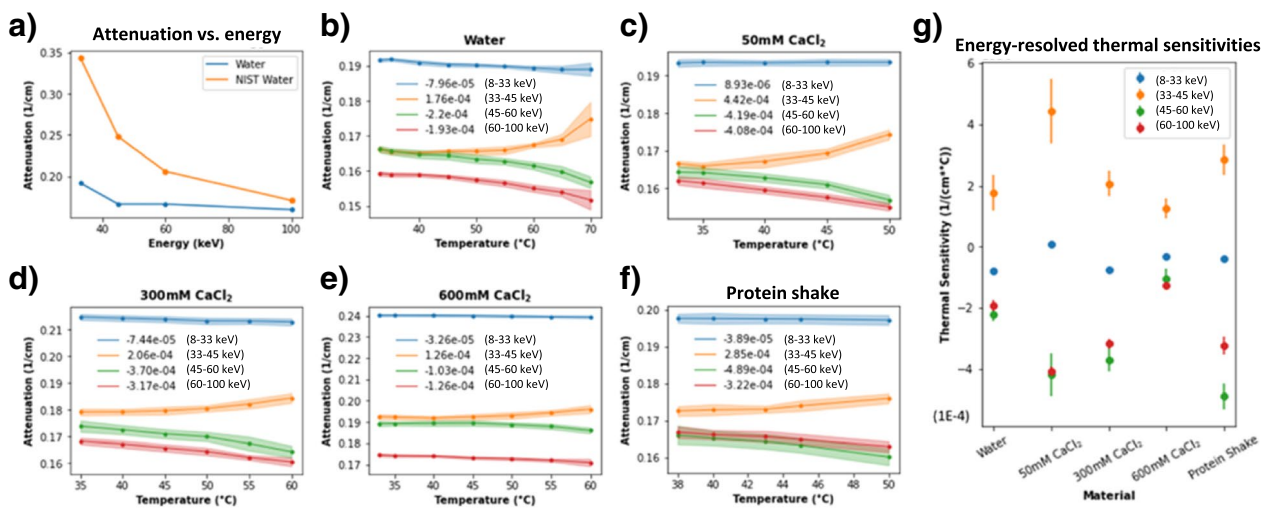

**Fig. 2** Graphs of attenuation data of all studied materials. **a** Attenuation vs energy plots for water compared to NIST values (we multiply the mass attenuations by the density of water, assumed to be 1.00 g/cm$_3$ at 33 °C). The end points of the energy bins (33, 45, 60, and 100 keV) were selected for the figure. Compton scattering of high energy photons accounts for the observed attenuation discrepancy; **b-f** Attenuation vs temperature plots for all materials. The positive trend in the 33-45 keV bin is due to the effects of temperature on Compton scattering at high energies. The legend indicates the slope of the regression line for the color-coded trend; **g** Scatterplot summary of thermal sensitivities with error bars. Observe the non-linear trend between thermal sensitivity and composition for 0, 50, 300, and 600 mmol/L solutions of $CaCl_2$

temperature drop. The system consists of an X-ray source (SourceRay SB-120-350, 75 μm focus) and an X-ray photon-counting detector (ADVACAM WidePIX1x5, Medipix3, 55 μm pitch, 256 × 1280 pixels). In our experiments, the source was operated at 100 kVp 100 μA with 0.1 mm copper filtration. The detector was set to the charge-summing mode with two thresholds for each acquisition. After 1 h of stabilization, projections were collected at 8 keV and 45 keV thresholds followed by the same number of projections at thresholds of 33 keV and 60 keV. All projections were captured within a 1.5 °C change of the digital thermometer reading.

Since the X-ray tube emits photons in a small-angle cone geometry, we cannot assume that all beam paths through the phantom are in parallel. Thus, a weak perspective method was used to compensate for beam divergence. This is illustrated in Fig. 1b and c. In the 2D projection after removal of a small proportion of unstable pixels (greater than 3 standard deviations from the average), we selected a horizontal LOI that spans the width of the phantom [17]. Using $x$ to denote position along the LOI, the difference between the line integral profiles of the phantom when it is filled with liquid and when it is empty was computed according to Eq. 4.

$$\Delta p(x) = p_f(x) - p_e(x) = L * \mu_{\text{material}}(x) = \ln\left(\frac{I_e(x)}{I_f(x)}\right) \quad (4)$$

where $f$ stands for filled, $e$ stands for empty, $I$ are the raw photon counts, and $L$ (254 mm) is the external side length of the square cross-section of the phantom. By taking the difference, the attenuation contribution of the phantom enclosure was eliminated. Finally, a sliding average over five pixels and a median filter over seven pixels were sequentially applied to remove noise from the profiles before the attenuation of the material is found in Eq. 5.

$$\mu_{material} = \frac{0.055}{L^2} \int \frac{\Delta p(x)}{1.23} dx \quad (5)$$

where the correction factor of 1.23 is the magnification, defined as the ratio of the distances from source to the phantom center (278.5 mm) and from source to detector (342.5 mm) and 55E-3 mm is the length of a detector pixel. The error in $\mu_{\text{material}}$ is theoretically no more than 3% compared to if it were measured with a parallel beam source. Note that our weak perspective method is rotation-invariant and uses all data points in the LOI to yield a high signal to noise ratio. The variance of all measurements was quantified by computing the LAC as the average of 10 adjacent LOI's.

To predict the temperature changes, we designed a neural network with an input layer of eight nodes, two hidden layers of four nodes, and an output layer of 1 node. The training examples were generated from the base material data. Shown in Eq. 6, the first four elements of the input are a material's LACs at some temperature and the last four are the LAC residuals due to heating of the material above the 33 °C baseline. The multiplicative factor of 100 was introduced to scale the residuals into a similar range as the baseline. The network architecture is displayed in Fig. 3a.

$$input = \begin{bmatrix} \mu_1(T, E_1) \\ \vdots \\ \mu_1(T, E_4) \\ [\mu_1(T, E_1) - \mu_1(T_0, E_1)] * 100 \\ \vdots \\ [\mu_1(T, E_4) - \mu_1(T_0, E_4)] * 100 \end{bmatrix} \quad (6)$$

In total, 333 unique training inputs representing a reasonable range of temperatures were generated for each of the three base materials where a small amount of Gaussian random noise was added to each input. The ReLU activation was used for all layers, mean squared error acted as the loss function, and stochastic gradient descent with a learning rate of 1E-5 was used as the optimizer. The dataset was split 80% for training and 20% for validation. The testing set consisted of data collected from 300 mmol/L aqueous $CaCl_2$, which is similar in composition to the base materials, and from a milk-based protein shake (30 g protein, 4 g carbohydrates, 2.5 g fat per 340 mL), which is organic and dissimilar to the base materials. The uncertainty of the temperature predictions is quantified by evaluating the network on the testing

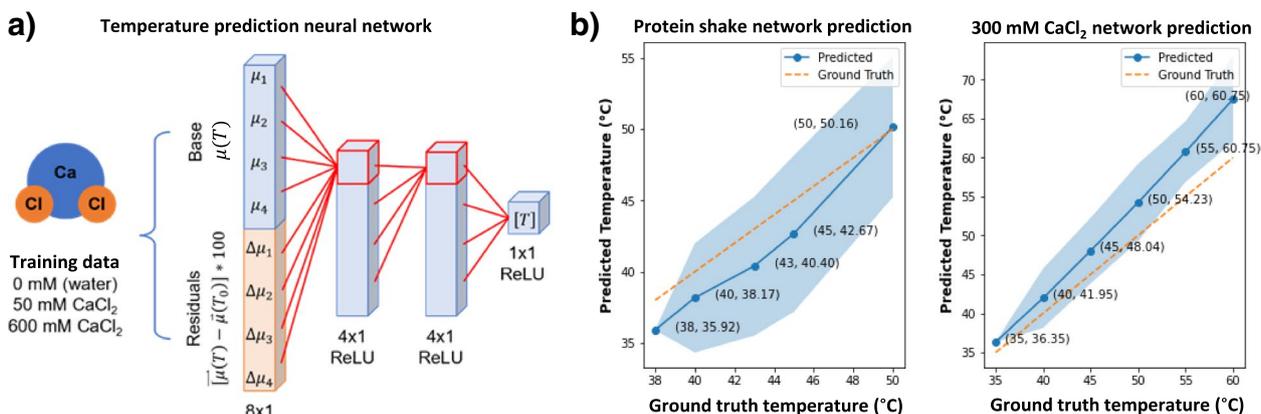

**Fig. 3** Summary of experiment results. **a** The fully connected neural network architecture used to non-linearly model the relationship between attenuation and temperature. The input to the network are the spectral attenuations of a material at a baseline temperature concatenated with the attenuation residuals due to heating; **b** Visualization for network performance for predicting temperature on 300 mmol/L $CaCl_2$ and a milk-based protein shake. The data points are labeled in the (xx, yy) format where xx is the predicted temperature and yy is the ground truth temperature synchronously measured with a digital thermometer. The 95%CI of temperature prediction is shaded. Data from the testing samples were not included in the training data

data with randomly generated Gaussian noise. This noise is distributed according to the variance in attenuation obtained from 10 LOI's in the corresponding projection. Hence, we realistically simulate the range of attenuation values that are measured in practice.

## Results and discussion

All the collected data is illustrated in Fig. 2 and the raw data and code are made openly available [18]. Figure 2a depicts the trend of X-ray LAC with increasing energy levels, which is generally expected. However, at low energies, our measured attenuation coefficients for water are lower than those reported by NIST [19]. This discrepancy is due to Compton scattering of high energy photons from our polychromatic source which were recorded as low energy photons.

Figure 2b-f show the relationship between attenuation and temperature, which is a negative trend in all except the 33-45 keV channel. The reduced attenuation of a material due to thermal expansion leads to two competing effects: fewer high energy (45-60 keV and 60-10 keV) photons are Compton scattered while more low energy (33-45 keV) photons pass through. It is hypothesized that the former phenomenon has a greater effect since it occurs over a wider energy range. Hence, the net effect is that attenuation is increased with increasing temperature in the 33-45 keV bin. Despite this effect, the data in the 33-45 keV channel is still informative and is incorporated into the network.

After 73 epochs of training, the MAE on the validation data smoothly converged from 43.13 °C to 3.40 °C. The network is evaluated by on the testing materials by taking a baseline scan and computing the residuals from heating in an identical fashion as described in Eq. 6 for the training data. On the testing set, the network achieves a MAE of 3.97 °C on 300 mmol/L $CaCl_2$ over a temperature range of 35 °C to 60 °C and an MAE of 1.80 °C on a milk-based protein shake over a temperature range of 38 °C to 50 °C. Note that 300 mmol/L $CaCl_2$ can be directly made from the bases (i.e., 50% water and 50% 600 mmol/L $CaCl_2$) while the protein shake must be indirectly modeled since it contains significant amounts of other substances. In both cases, the network is highly accurate. These results are displayed in Fig. 3b and c.

## Conclusions

In future studies, an active temperature measure (as opposed to passive cooling) could be used to ensure better thermal accuracy of the data points. A better calibrated PCD and increased source filtration can also reduce the adverse effects of fluorescence escape and beam hardening effects respectively [20, 21].

Additionally, more material bases can be incorporated for the neural network to cover more material types and better neural networks can be designed to improve temperature prediction. Furthermore, tomographic PCCT on human tissue samples are necessary before in vivo studies can be planned. For preclinical evaluation, mouse experiments can be used to compare the efficacy of thermal ablation using classical approaches (e.g., thermistors) and the novel PCCT thermometry imaging presented in this letter. Clearly, PCCT thermometry will offer a thermal dimension to a spectral CT volume and may potentially bring new diagnostic and therapeutic tools to clinical practice. Furthermore, the idea of using material decomposition to improve thermometry may also be applied to phase contrast X-ray thermometry, which has been shown to be capable of volumetric thermal visualization [22].

In this study, we demonstrate a data driven PCCT thermometry algorithm that can accurately predict the temperature of unknown materials given spectrally resolved LACs of a set of known, base materials at various temperatures. This is an important result toward surgical translation as it presents a solution for handling variability in tissue property without direct calibration to the tissue in vivo.

**Abbreviations**
| | |
|---|---|
| CT | Computed tomography |
| US | Ultrasound |
| MRI | Magnetic resonance imaging |
| PCCT | Photon counting computed tomography |
| LAC | Linear attenuation coefficient |
| HU | Hounsfield units |
| LOI | Line of interest |
| MAE | Mean absolute error |

**Acknowledgements**
We thank the anonymous reviewers very much for their constructive comments.

**Authors' contributions**
NW developed the neural network for temperature prediction, led the experiments, and drafted the manuscript; MZ contributed to the experimental design and data collection and revised the manuscript; PH built the phantom system for the experiments. All authors read and approved the final manuscript.

**Funding**
This work is supported by the Johns Hopkins University Leong Research Award for Undergraduates.

**Availability of data and materials**
Data and code underlying the results presented in this paper are available in Dataset 1, Ref. [18] or may be requested from the authors.

**Declarations**

**Competing interests**
The authors declare no conflicts of interest.

### References

1. Fani F, Schena E, Saccomandi P, Silvestri S (2014) CT-based thermometry: An overview. Int J Hyperthermia 30(4):219-227. https://doi.org/10.3109/02656736.2014.922221
2. Brace C (2011) Thermal tumor ablation in clinical use. IEEE Pulse 2(5):28-38. https://doi.org/10.1109/MPUL.2011.942603
3. Alkhorayef M, Mahmoud MZ, Alzimami KS, Sulieman A, Fagiri MA (2015) High-intensity focused ultrasound (HIFU) in localized prostate cancer treatment. Pol J Radiol 80:131-141. https://doi.org/10.12659/PJR.892341
4. Saccomandi P, Frauenfelder G, Massaroni C, Caponera MA, Polimadei A, Taffoni F et al (2016) Temperature monitoring during radiofrequency ablation of liver: In vivo trials. Paper presented at the 2016 38th annual international conference of the IEEE engineering in medicine and biology society. IEEE, Orlando https://doi.org/10.1109/EMBC.2016.7590710
5. Saccomandi P, Schena E, Silvestri S (2013) Techniques for temperature monitoring during laser-induced thermotherapy: An overview. Int J Hyperthermia 29(7):609-619. https://doi.org/10.3109/02656736.2013.832411
6. Winter L, Oberacker E, Paul K, Ji YY, Oezerdem C, Ghadjar P et al (2016) Magnetic resonance thermometry: Methodology, pitfalls and practical solutions. Int J Hyperthermia 32(1):63-75. https://doi.org/10.3109/02656736.2015.1108462
7. Ebbini ES, Simon C, Liu DL (2018) Real-time ultrasound thermography and thermometry [life sciences]. IEEE Signal Process Mag 35(2):166-174. https://doi.org/10.1109/MSP.2017.2773338
8. Wang G, Yu HY (2013) The meaning of interior tomography. Phys Med Biol 58(16):R161-R186. https://doi.org/10.1088/0031-9155/58/16/R161
9. Wang G (2016) A perspective on deep imaging. IEEE Access 4:8914-8924. https://doi.org/10.1109/ACCESS.2016.2624938
10. Taguchi K, Polster C, Segars WP, Aygun N, Stierstorfer K (2022) Model-based pulse pileup and charge sharing compensation for photon counting detectors: A simulation study. Med Phys 49(8):5038-5051. https://doi.org/10.1002/mp.15779
11. Willemink MJ, Persson M, Pourmorteza A, Pelc NJ, Fleischmann D (2018) Photon-counting CT: Technical principles and clinical prospects. Radiology 289(2):293-312. https://doi.org/10.1148/radiol.2018172656
12. Homolka P, Gahleitner A, Nowotny R (2002) Temperature dependence of HU values for various water equivalent phantom materials. Phys Med Biol 47(16):2917-2923. https://doi.org/10.1088/0031-9155/47/16/307
13. Heinrich A, Schenkl S, Buckreus D, Güttler FV, Teichgräber UKM (2022) CT-based thermometry with virtual monoenergetic images by dual-energy of fat, muscle and bone using FBP, iterative and deep learning-based reconstruction. Eur Radiol 32(1):424-431. https://doi.org/10.1007/s00330-021-08206-z
14. Pandeya GD, Klaessens JHGM, Greuter MJW, Schmidt B, Flohr T, van Hillegersberg R et al (2011) Feasibility of computed tomography based thermometry during interstitial laser heating in bovine liver. Eur Radiol 21(8):1733-1738. https://doi.org/10.1007/s00330-011-2106-6
15. Li ZB, Leng S, Yu LF, Yu ZC, McCollough CH (2015) Image-based material decomposition with a general volume constraint for photon-counting CT. Proc SPIE Int Soc Opt Eng 9412:94120T. https://doi.org/10.1117/12.2082069
16. Yang QS, Cong WX, Wang G (2015) Material decomposition with dual energy CT. Paper presented at the 2015 41st annual northeast biomedical engineering conference. IEEE, Troy
17. Li MZ, Lowe C, Butler A, Butler P, Wang G (2022) Motion correction via locally linear embedding for helical photon-counting CT. arXiv:2204.02490. https://doi.org/10.1117/12.2646714
18. Dataset 1. https://github.com/nathanwangai/pcct_thermometry. Accessed 15 Oct 2022.
19. NIST: X-Ray Mass Attenuation Coefficients - Water, Liquid. https://physics.nist.gov/PhysRefData/XrayMassCoef/ComTab/water.html. Accessed 15 Oct 2022.
20. Li MZ, Rundle DS, Wang G (2020) X-ray photon-counting data correction through deep learning. arXiv:2007.03119
21. Li MZ, Fan FL, Cong WX, Wang G (2021) EM estimation of the X-ray spectrum with a genetically optimized step-wedge phantom. Front Phys 9:678171. https://doi.org/10.3389/fphy.2021.678171
22. Yoneyama A, Iizuka A, Fujii T, Hyodo K, Hayakawa J (2018) Three-dimensional X-ray thermography using phase-contrast imaging. Sci Rep 8(1):12674. https://doi.org/10.1038/s41598-018-30443-4

## Publisher's Note

Springer Nature remains neutral with regard to jurisdictional claims in published maps and institutional affiliations.